\documentstyle[prd,aps,epsf]{revtex}
\begin{document}
\draft
\twocolumn[\hsize\textwidth\columnwidth\hsize\csname@twocolumnfalse\endcsname
\title{Primordial Magnetic Fields from Dark Energy}
\author{Da-Shin Lee~\cite{dashin}}
\address{Department of Physics, National Dong Hwa University, Hua-Lien, Taiwan}
\author{Wolung Lee~\cite{wolung} and Kin-Wang Ng~\cite{kin}}
\address{Institute of Physics, Academia Sinica, Taipei, Taiwan}
\maketitle

\begin{abstract}
Evidences indicate that the dark energy constitutes about two thirds
of the critical density of the universe. If the dark energy is an
evolving pseudo scalar field that couples to electromagnetism,
a cosmic magnetic seed field can  be produced via spinoidal instability
during the formation of large-scale structures.
\end{abstract}

\vspace{2pc}]

Recent astrophysical and cosmological observations such as dynamical mass,
Type Ia supernovae (SNe), gravitational lensing,
and cosmic microwave background (CMB) anisotropies, concordantly
prevail a spatially flat universe containing a mixture of matter and
a dominant smooth component with effective negative pressure~\cite{wang}.
The simplest possibility for this component is a cosmological constant.
A dynamical variation calls for the existence of dark energy
whose equation of state approaches that of the cosmological constant at recent
epochs. Many possibilities have been proposed to explain for the dark energy.
Most of the dark energy models involve the dynamical evolution
of classical scalar fields or quintessence (Q). For a Q-model,
the dynamics is governed by the scalar field potential such that
the vacuum energy becomes dynamically important only at recent
epochs and gives rise to an effective cosmological constant today.
So far, many different kinds of scalar field potentials have been proposed.
They include pseudo Nambu-Goldstone boson (PNGB),
inverse power law, exponential,
tracking oscillating, and others~\cite{waga}.
Upcoming observations will measure the
equation of state so as to discriminate between these models and
distinguish them from the cosmological constant~\cite{huterer}.

Since the scalar potential $V(\phi)$ of the Q field is scarcely known,
it is convenient to discuss the evolution of $\phi$
through the equation of state, $p_\phi=w_\phi\rho_\phi$. Physically,
$1\ge w_\phi\ge -1$, where the latter equality holds for a pure vacuum state.
Lately some progress has been made in constraining the
behavior of Q fields from observational data.
A combined large scale structure (LSS), SNe, and CMB analysis has set an upper
limit on Q-models with a constant $w_\phi < -0.7$ ~\cite{bond},
and a more recent analysis of CMB observations gives
$w_\phi=-0.82^{+0.14}_{-0.11}$~\cite{bac}. Furthermore,
the SNe data and measurements of the position of the acoustic peaks in the CMB
anisotropy spectrum have been used to put a constraint on the present
$w_\phi^0 \le -0.96$~\cite{cope}.The apparent brightness of the farthest
SN observed to date, SN1997ff at redshift $z\sim 1.7$, is consistent with
that expected in the decelerating phase of the flat $\Lambda$CDM model with
$\Omega_\Lambda \simeq 0.7$~\cite{riess}, inferring $w_\phi= -1$ for $z<1.7$.
In addition, several attempts have been made to test different
Q-models~\cite{many}. Nevertheless, it is primitive to differentiate between
the variations, and the reconstruction of $V(\phi)$
would require next-generation observations.

Although the physical state of the dark energy can be probed through
its gravitational effects on the cosmological evolution, it is important
in fundamental physics to understand whether the quintessence is
a nearly massless, slowing rolling scalar field. It has been pointed out
by Carroll that the existence of  an approximate global symmetry would allow a coupling of the Q field, $\phi$,  to
the pseudoscalar $F_{\mu\nu}{\tilde F}^{\mu\nu}$ of
electromagnetism, which provides a potential observable in polarization studies of distant radio sources~\cite{carr}.
As long as an ultralight $\phi$ field couples to photon where  for a slow-roll condition, the mass $m_\phi$ is comparable to $H_0$, it is conceivable to
have very long-wavelength electromagnetic fields  generated via spinodal
instabilities from the dynamics of $\phi$ as a possible source of seed magnetic fields for the galactic dynamo.

As we know,  the  issue of the origin of the observed cluster and galactic magnetic fields
of about a few $\mu {\rm G}$~\cite{kron} remains a puzzle~\cite{pmf}.
These magnetic fields could have been resulted from the amplification
of a seed field of $B_{seed} \sim 10^{-23} {\rm G}$ on a comoving scale
larger than Mpc via the so-called galactic dynamo effect.
A number of scenarios have been proposed
for generating seed fields in the early universe,
mainly relying on non-equilibrium conditions such as inflation,
the electroweak and the QCD phase transitions. After the phase transitions,
the universe became a highly conducting plasma so that the magnetic flux
which existed is frozen in, and the ratio of the magnetic energy density
and the thermal background, $\rho_B/\rho_\gamma$, remains constant thereafter.
The required $B_{seed}$ translates into $\rho_B\simeq 10^{-34}\rho_\gamma$.
However, it turns out that the generated fields in these models are too small,
except in somewhat contrived cases, to be of cosmological interest.
In an equilibrium condition, a large damping term induced by the high
plasma conductivity suppresses significantly any
electromagnetic field fluctuations~\cite{turner,gio}.

In this {\it Letter}, we will investigate the implication of the electromagnetic
coupling of the evolving  $\phi$ field to the origin of the primordial magnetic field (PMF). Then, we will provide  a mechanism in which the PMF can be generated
via the $\phi$-photon resonant conversion during the LSS formation.

\begin{figure}
\leavevmode
\hbox{
\epsfxsize=3in
\epsffile{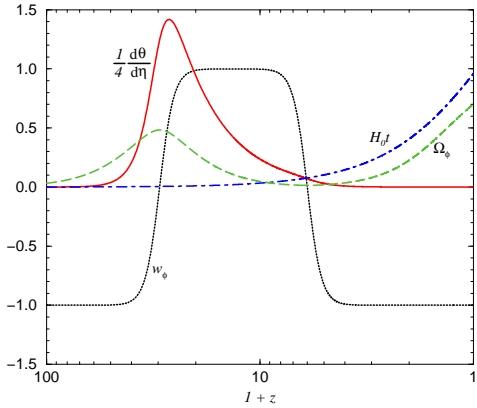}}
\caption{The quantities $d\theta/d\eta$, $w_\phi$, $H_0 t$, and $\Omega_\phi$
as a function of redshift. Note that $d\theta/d\eta$ is drawn 4 times smaller.}
\label{fig1}
\end{figure}

Here we consider the $\phi$-photon coupling,
\begin{equation}
L_{\phi \gamma} = \frac{c}{f} \phi\ \epsilon^{\alpha\beta\mu\nu}
            F_{\alpha\beta} F_{\mu\nu},
\end{equation}
where $F_{\mu\nu}=\partial_\mu A_\nu - \partial_\nu A_\mu$, $f$ is a mass
scale, and $c$ is a coupling constant which we treat as a free parameter.
For the present consideration, we pick $f$ equal to the reduced
Planck mass $M_p\equiv (8\pi G)^{-1/2}$.
We are thus led to study the cosmological evolution of the
$\phi$-photon system in a flat universe. The effective action is
\begin{eqnarray}
S= &&S_M+ \int d^4x {\sqrt g} \left[ -\frac{R}{16\pi G}
   - {1\over4} g^{\alpha\mu} g^{\beta\nu} F_{\alpha\beta} F_{\mu\nu}
   \right. \nonumber \\
  &&\left. -{1\over2}g^{\mu\nu}\partial_\mu\phi\partial_\nu\phi
               - V(\phi) + \frac{1}{\sqrt g} L_{\phi\gamma} \right],
\label{action}
\end{eqnarray}
where the signature is $(-+++)$ and $S_M$ is the classical action for matter.
We assume that the universe today has matter $\Omega_M^0= 0.3$ and
quintessence $\Omega_\phi^0=0.7$,  and define an
$\Omega_\phi$-weighted average\cite{wang}
\begin{equation}
\langle w_\phi \rangle = \int_{\eta_{ls}}^{\eta_0}
\Omega_\phi(\eta) w_\phi(\eta) d\eta
\times \left( \int_{\eta_{ls}}^{\eta_0} \Omega_\phi(\eta) d\eta \right)^{-1},
\end{equation}
where $\eta_0$ and $\eta_{ls}$ are respectively the conformal time today and
at the last scattering, defined by $\eta=H_0\int dt a^{-1}(t)$
with the scale factor $a$ and the Hubble constant $H_0$.
Assuming a spatially homogeneous $\phi$ field,
the equation of motion is given by
\begin{equation}
\frac{d^2\theta}{d\eta^2}+\frac{2}{a}\frac{da}{d\eta}\frac{d\theta}{d\eta}
+\frac{a^2}{H_0^2 M_p^2}\frac{\partial V(\theta)}{\partial\theta} 
-\frac{c}{a^2 H_0^2 M_p^2}\langle F \tilde F \rangle =0,
\label{eom}
\end{equation}
where $\theta=\phi/M_p$. The last term of Eq.~(\ref{eom}) is the back reaction
from the produced magnetic fields. Later we will show that this term 
is too small to have any effect on the evolution of the $\phi$ field
when we consider the photon production.
For now we omit this term and thus 
the evolution of the cosmic background is governed by
\begin{eqnarray}
\frac{d{\bar\rho}_\phi}{d\eta}&=& -3ah \left(1+w_\phi\right){\bar\rho}_\phi, \\
\frac{dh}{d\eta}&=& -{3\over2}ah^2- {1\over2}aw_\phi{\bar\rho}_\phi,
\end{eqnarray}
where $h=H/H_0$, ${\bar\rho}_\phi=\rho_\phi/(M_p H_0)^2$, and we
have used ${\dot\phi}^2=(1+w_\phi)\rho_\phi$ and
$V(\phi)=(1-w_\phi)\rho_\phi/2$. We have numerically solved the
background equations by proposing a simple square-wave form for
$w_\phi$ as shown in Fig.~\ref{fig1}. In order to satisfy the
above-mentioned observational constraints on $w_\phi$, we have
chosen $w_\phi= -1$ for $z<2$, and a width of the square-wave such
that $\langle w_\phi \rangle\simeq -0.7$. Note that at the present
time $\Omega_\phi={\bar\rho}_\phi/(3h^2)=0.7$ and $H_0 t=0.95$. We
have also plotted $d\theta/d\eta=a{\sqrt{(1+w_\phi){\bar\rho}_\phi}}$. 

In the original PNGB models for dark energy~\cite{pngb}, the well-known
PNGB cosine potential was used. This results in that $w_\phi>0$ at the
present time, which is disfavored by the observations.
In fact, the square-wave equation of state is anticipated and quite general 
in the class of Q-models using PNGB fields incorporated 
with quantum corrections to the cosine potential~\cite{corm}. 
In the PNGB models, the nonperturbatively large
quantum fluctuations driven by spinoidal instabilities   strongly
dissipate the oscillation of the scalar field expectation value,
leading  to the field  oscillation to become overdamped, and
finally the field expectation value relaxes  to the minimum of the
effective action. However,  we would like to emphasize here that
the spinodal instabilities that we will discuss later to  drive
the amplitude fluctuations of long-wavelength magnetic fields to
grow  is a general phenomenon which  is {\it not} restricted to
such a square-wave equation of state. In fact, in the present
consideration, the spinoidal instabilities responsible for the
generation of the PMF occur during the formation of LSS for $z\sim
10-2$ as long as during which  the equation of state of the scalar
field deviates from $w_\phi=-1$. After $z\sim 2$, the
quintessential potential dominates and behaves like a cosmological
constant with $w\sim -1$ that ceases the photon production.
Several quintessence models have been proposed to generate   the
equation of state of the scalar field that has  the
above-mentioned features. In Ref.~\cite{skordis}, a potential of
the form $V(\phi)=V_p(\phi) e^{-\lambda\phi}$, where $V_p(\phi)$
is a polynomial of $\phi$ and $\lambda$ is a parameter, has been
invented to introduce a local minimum in the exponential such that
the field gets trapped into it. After the field gets trapped it
starts behaving like a cosmological constant and the universe
eventually enters an era of accelerated expansion. With suitable
parameters, they showed that the equation of state stays near the
value of   about $0.3$ for high redshifts and abruptly decreases
to $-1$ for $z\alt 3$. This desired accelerated expansion can be
also provided by a different $V_p(\phi)$ arising from the
interaction between branes in extra-dimension
models~\cite{skordis}, or having an oscillating term in the
tracking oscillating model by Dodelson {\it et al.} in
Ref.~\cite{waga}. In $k$-essence models with a modified kinetic
term for the scalar field, the equation of state suddenly drops to
the value of about $-1$ at $z\sim 2$~\cite{armen}. In summary, we
have reviewed some quintessence models which have the desired
evolution of the scalar field. With or without further tuning of
the model parameters, they can be used to provide a cosmological
background for the generation of the PMF during the LSS formation
with the photon production ending at $z\sim 3$. However, for our
purpose to illustrate how  the PMF  can be generated from  the
quintessence dynamics,  we will use this simple, generic model in
Fig.~\ref{fig1} that satisfies all existing observational
constraints to carry out the calculation.

The most stringent limit on the $\phi$-photon coupling comes from
the cooling via the Primakoff conversion of horizontal branch (HB)
stars in globular clusters~\cite{raf}, $c/f < 1.5 \times 10^{-11}
{\rm GeV^{-1}}$, which gives $c < 3.7\times 10^7$. If $\phi$
carries a mass $m_\phi \sim H_0$, it would decay into two photons
with a width, $\Gamma = c^2 m_\phi^3/(4\pi f^2)$. Hence the
lifetime of $\phi$ is much longer than the age of the universe.
There is a very nice limit on $c$ coming from the rotation of the
plane of polarization of light from distant radio sources, $c <
3\times 10^{-2}/|\Delta\theta|$, where $\Delta\theta$ is the
change in $\theta$ between $z=0.425$ and today~\cite{carr}. Since
we have taken $w_\phi=-1$ (i.e., $\Delta\theta=0$) for $z<2$, this
limit does not apply to our case at all. From the theoretical
point of view, it has been argued that because of radiative
corrections $c$ must be infinitesimally small to keep the $\phi$
field light~\cite{pec}. However, in this study we would like to
find out the value of the $\phi$-photon coupling needed in order to
produce the large PMF over large correlation length scales if
we assume that the generation of the PMF is solely due to the
quintessence dynamics during the LSS formation.

From the action~(\ref{action}), the comoving magnetic field in the
comoving coordinates $(\tau,{\bf x})$ with $\tau=\eta/H_0$ satisfies
\begin{eqnarray}
\left({\bf \nabla}^2-\frac{\partial^2}{\partial\tau^2}\right) {\bf B}=&&
\sigma a \left[\frac{\partial {\bf B}}{\partial\tau} -
{\bf \nabla} \times ( {\bf v} \times {\bf B}) \right] \nonumber \\
&& + 4 c \frac{d\theta}{d\tau} {\bf \nabla} \times {\bf B},
\label{fulleq}
\end{eqnarray}
where ${\bf \nabla}\equiv \partial/\partial{\bf x}$, $\sigma$ is
the plasma conductivity, and ${\bf v}=d{\bf x}/d\tau$ is the
peculiar plasma velocity. After hydrogen recombination, the
residual ionization keeps the conductivity high, $\sigma/H\sim
10^{22}(T/{\rm eV})^{-3/2} \gg 1$. As a result, the $\sigma$ term
under the assumption that ${\bf v}=0$ damps out any growth of the
$\bf B$ field on scales above $\sim$ A.U.~\cite{turner}. But this
assumption may not hold when the universe has entered the
non-linear regime. It could be understood from the recent study in
Ref.~\cite{kul} where the battery mechanism has been proposed
 as a source for generating a small initial
magnetic field during LSS formation.

From the theory of small fluctuations, $\bf v$ will grow with density
perturbations. Velocity flows grow rapidly after
extreme nonlinearities develop in the cosmic fluid.
Therefore, the authors in Ref.~\cite{kul} performed numerical simulations 
by adding the battery equation to the hydrodynamic code for structure 
formation, and
showed that magnetic fields are built up in regions about to collapse
into galaxies. The non-linear turbulence effect plays an important role in  
generating and subsequently  amplifying  the created  PMF which  overcomes
the  damping mechanism due to the high plasma conductivity.
The numerical simulations showed that the rms value for
${\bf \nabla}\cdot{\bf v} \sim {\bf \nabla}\times{\bf v}$ is about
$10^{-16} s^{-1}$ on comoving scale of Mpc at $z \sim 3$~\cite{kul}.
This is just the inverse of the time scale $\tau_{LSS}$ for LSS formation.
When a Mpc-scale magnetic field with magnitude $B$ is simultaneously
created, and as long as  the growth rate ${\partial B}/{\partial\tau} \sim B/\tau_{LSS}$ is
comparable to the twisting term $|{\bf \nabla} \times ( {\bf v}
\times {\bf B})| \sim B{\bf \nabla}\cdot{\bf v}$,  the $\sigma$ term would be
 significantly reduced and the high
plasma conductivity is no longer a hindrance to the growth of
large-scale magnetic fields. Here we will show that the temporal variation of $\phi$
can also be a generating source for large-scale magnetic fields. However, we will not
pursue numerical simulations similar to Ref.~\cite{kul},
combining  their battery source with  the quintessence dynamics.
As a first step, we will simply omit the $\sigma$ term and solve for the
photon equation self-consistently,
while following closely their results and estimating the effect of
the magneto-hydrodynamical dissipation.
Of course, it would be very interesting to carry out
direct numerical simulations by coupling the $\phi$-photon system
in Eq.~(\ref{fulleq}) to
the hydrodynamic code to make a map of the created magnetic fields.

Now we write ${\bf B}={\bf \nabla}\times{\bf A}_T$ and
decompose the transverse field ${\bf A}_T(\tau,{\bf x})$ into Fourier modes,
\begin{equation}
{\bf A}_T =
 \int \frac{d^3{\bf k}}{\sqrt{2(2\pi)^3 k}}
 \left[ e^{i{\bf k}\cdot {\bf x}} \sum_{\lambda=\pm} b_{\lambda {\bf k}}
 V_{\lambda {\bf k}}(\tau) {\bf \epsilon}_{\lambda {\bf k}}
 + {\rm h.c.} \right],
\end{equation}
where $b_{\pm {\bf k}}$ are destruction operators,
and ${\bf \epsilon}_{\pm {\bf k}}$ are circular polarization unit vectors.
Then, defining $q=k/H_0$, the mode equations are
\begin{equation}
\frac{d^2}{d\eta^2}V_{\pm q}+
\left(q^2\mp 4cq\frac{d\theta}{d\eta}\right)V_{\pm q}=0,
\label{modeeq}
\end{equation}
with initial conditions at early epoch given by
\begin{equation}
V_{\pm q}=1,\quad \frac{dV_{\pm q}}{d\eta}=-iq.
\end{equation}
Hence, the comoving energy density of the magnetic field is given by
$\rho_B=\langle B^2\rangle/8\pi=\int dq (d\rho_B/dq)$ with
\begin{equation}
\frac{d\rho_B}{dq}= \frac{H_0^4}{32\pi^3}
q^3 \coth \left[ \frac{qH_0}{2 T_0} \right]
\sum_{\lambda=\pm}|V_{\lambda q}|^2,
\end{equation}
where the coth term is the Bose-Einstein enhancement factor due to the
presence of the CMB with current temperature $T_0$ and energy
density $\rho_{\gamma}=\pi^2T_0^4/15$. The nonlinear growth in
$V_{+ q}$ can be understood by treating the background field
solution ${d\theta}/{d\eta} >0 $ in Eq.~(\ref{modeeq}) as a
constant. For small fluctuations of the field, the long-wavelength
modes of  $V_{+ q}$ , where $q$ lies within the unstable band, $q
< 4c({d\theta}/{d\eta})$, in fact  ''see'' the  inverted harmonic
oscillators that cause the amplitude fluctuations to grow
exponentially, while the evolution of $V_{-q}$ modes is purely
oscillating  without any growth. The growth of these unstable
modes is driven by spinodal instabilities~\cite{dan}. Now we can
write the approximate solution for the growing modes to
Eq.~(\ref{modeeq}) as $V_{+ q}\propto e^{\alpha \eta}$ where the
exponent $\alpha= \sqrt{4cq({d\theta}/{d\eta}) - q^2}$ provided
that $q$ is in the unstable band.  The maximum value of $\alpha$
is $\alpha_{\rm max} = 2c({d\theta}/{d\eta})$ at
$q=2c({d\theta}/{d\eta})$. Since the maximum  growth  for the
amplitude fluctuations occurs at $q=2c({d\theta}/{d\eta})$, in
order to produce the large-scale  correlated magnetic fields on
$10$ Mpc scales with the fastest growth, one can  estimate  the
order-of-magnitude of the coupling $c$, which is  about $ 10^2 $,
using the fact that ${d\theta}/{d\eta} \simeq 0.8 $ at redshift
$z=10$ after which the universe was in the nonlinear regime and
based on our previous argument that the  high plasma conductivity
effect which could  damp out the produced magnetic fields can be
consistently  ignored. The self-consistency condition to reduce
significantly  this high conductivity effect  can be justified in
the sense  that for the exponentially growing modes that we are
interested in,   ${\partial B}/{\partial\tau} \sim \alpha_{\rm
max}H_0 B \sim (10^{-16} s^{-1}) B $ with the coupling $c \simeq
100$ and  ${d\theta}/{d\eta} \simeq 1 $, which is of the same
order of the twisting term $|{\bf \nabla} \times ( {\bf v} \times
{\bf B})| \sim B{\bf \nabla}\cdot{\bf v}$ during the LSS formation
obtained from the numerical simulations\cite{kul}.

\begin{figure}
\leavevmode
\hbox{
\epsfxsize=3in
\epsffile{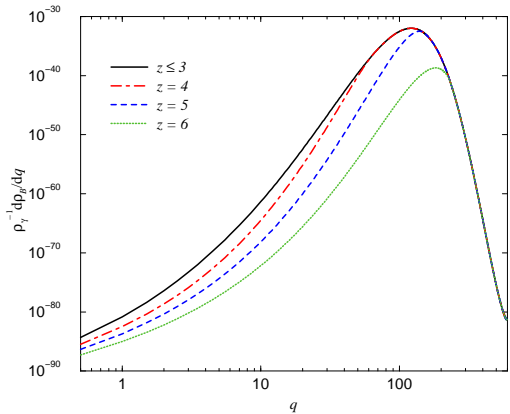}}
\caption{Ratios of the spectral magnetic energy density to the present CMB energy
density at various redshifts. The present wavelength of the magnetic field
is given by $2\pi/(qH_0)$.}
\label{fig2}
\end{figure}

As you will see in our numerical analysis, the spinodal
instability provides a robust mechanism  to generate the
long-wavelength fluctuations where the created magnetic fields
correspond to coherent collective behavior in which the fields
correlate themselves over a large distance of order of 10 Mpc.
However, since ${d\theta}/{d\eta}$ is actually time dependent, the
modes will grow only for a period of time before they move out of
the unstable band. In the end, the growth of the long-wavelength
modes will be shut off completely when the bandwidth of the
unstable band vanishes as ${d\theta}/{d\eta}$ reaches zero at
redshift $z\sim 4$. We have numerically solved the mode
equations~(\ref{modeeq}) using $c=130$ and the background solution
as shown in Fig.~\ref{fig1}, and plotted the ratio
$(d\rho_B/dq)/\rho_\gamma$ in Fig.~\ref{fig2}. Although photons
are being produced as the scalar field starts rolling at $z\sim
60$, we have counted the photons produced only after $z=10$ when
the universe has presumably entered the non-linear regime. The
result shows that a sufficiently large seed magnetic field of 10
Mpc scale  has been produced before $z\sim 4$. 
Moreover, we notice that when $c\sim 130$ the
spinoidal instability gives rise to magnetic seed fields of the
right magnitude and length scale. If we decrease the value of $c$,
the peaks of the curves in Fig.~\ref{fig2} will shift to the lower
left-hand corner. 

The magnetic energy density can be approximately  obtained  from
Fig.~\ref{fig2} and is about $\Omega_B\simeq 10^{-33} \ll \Omega_{\phi}$. 
Therefore, we anticipate that the back reaction effect from the produced 
magnetic fields on the scalar field evolution equation is negligible. 
This can be shown by evaluating the last term of Eq.~(\ref{eom}) with
\begin{equation}
\langle F \tilde{F} \rangle=
\frac{H_0^4}{\pi^2} \int dq q^2 \coth \left[ \frac{qH_0}{2 T_0} \right]
\frac{d}{d\eta} \left( |V_{+q}|^2- |V_{-q}|^2 \right),
\end{equation}
which is found to be extremely small compared to the other terms in 
Eq.~(\ref{eom}) when the photons are being profusely produced.

Although the value of $c=130$ is well below the
HB limit, undoubtedly it is much larger than the theoretical
expectation. However, as suggested in Ref.~\cite{carr}, an
unsuppressed $\phi$-photon coupling may arise in higher
dimensional theories. One possible way to reduce the value of the
coupling  and at the same time to produce the sufficiently large
PMF on large correlation length scales during the LSS
formation is to combine the mechanism we proposed here with the
battery source in Ref.~\cite{kul}.
Also, note that the growth rate is controlled by the exponent 
$\alpha \sim 2c(d\theta/d\eta)$.
As such, it may be possible to reduce the value of $c$ by 
having a large $d\theta/d\eta$.
In the case of scalar quintessence, we have seen in Fig.~\ref{fig1} that
$d\theta/d\eta < 0.8$ for $z<10$. We cannot further increase
$d\theta/d\eta$ since $w_\phi$ has already reached the maximum value. 
Perhaps, in the $k$-essence models, in which
the kinetic term is modified to $K(\phi){\dot\phi}^2/2$ and we have
${\dot\phi}^2=(1+w_\phi)\rho_\phi/K(\phi)$, we may make 
$d\theta/d\eta$ much larger than one by tuning the prefactor $K(\phi)$. 
This possibility is under investigation.

In conclusion, we have made an effort to link the dark-energy
problem to a solution to another important problem in cosmology,
namely, the generation of primordial magnetic fields. So far, we
can only probe the equation of state of the dark energy through
its cosmological effects. Undoubtedly, it is extremely important
to have any clue about the nature of the dark energy. Here we have
studied the possibility of generating the PMF on Mpc scales during
the LSS formation by coupling the electromagnetic field to the
evolving  scalar field that accounts for the dark energy dynamics.
The high conductivity effect due to residual ionization after
hydrogen recombination can be argued to be significantly reduced
as a result of the existence of the cosmic flow with nonlinear,
twisting peculiar velocity to avoid a hindrance to the growth of
the magnetic fields. We have shown that the nonlinear instability
that drives the rapid growth of magnetic fields is of spinodal
instability where the long-wavelength modes about the Mpc scales
evolve as being the inverted harmonic oscillators and the
amplitude fluctuations begin to grow up to a time at which the
scalar field velocity approaches zero at redshift $z\sim 4$. We
found  that when $c\sim 130$ the spinoidal instability gives rise
to magnetic seed fields of the right magnitude and length scale.
This strong coupling is well below the HB limit, but much larger
than the theoretical expectation. 

Here we have used a standard scalar quintessence and a simple equation of
state as a working model. It is interesting to extend the idea to non-standard
scalar models such as $k$-essence to see if the required $\phi$-photon
coupling can be reduced to a smaller value with the specific equation of state.
On the other hand, new laboratory searches for the coupling of photon to 
pseudo-scalar and more ingenious astrophysical or cosmological limits on the 
coupling would be needed. In addition, future CMB and SNe experiments would 
put a strong constraint on the quintessential equation of state at 
high-redshifts. In particular,  combining  the
battery mechanism in  Ref.~\cite{kul}  with the  $\phi$-photon
coupling which provides an alternative robust mechanism to
generate the PMF in the hydrodynamical simulations would be
a very interesting subject to tackle.

We would like to thank Daniel  Boyanovsky and Hector de Vega for many interesting discussions and suggestions.
This work of D.S.L. (W.L. and K.W.N)  was  supported in part  from  National Science Council,
ROC under the Grant NSC 90-2112-M-259-011 (NSC 89-2112-M-001-060).

\end{document}